\documentclass[preprint,12pt]{elsarticle}

\usepackage{amssymb}
\usepackage{graphicx}
\usepackage{amsmath}
\usepackage{bm}
\usepackage{CJK}
\usepackage{url}
\usepackage{relsize}
\usepackage{float}      % for putting [H] condition to figures etc.
\usepackage[linesnumbered, ruled, vlined]{algorithm2e}
\usepackage[symbol]{footmisc}
\usepackage{listings}
\journal{Applied Mathematics and Computation}

\begin{document}

\begin{frontmatter}



\title{ZERNIPAX: A Fast and Accurate Zernike Polynomial Calculator in Python}


\author[inst1]{Yigit Gunsur Elmacioglu}
\author[inst2]{Rory Conlin\fnref{prev1}}
\author[inst3]{Daniel W. Dudt\fnref{prev1}}
\author[inst1]{Dario Panici}
\author[inst1,inst4]{Egemen Kolemen\corref{cor1}}

\fntext[prev1]{previously at Princeton University}

\cortext[cor1]{ekolemen@princeton.edu, +1  609-258-9209, D302D Engineering Quadrangle, Priceton, New Jersey, USA}
\affiliation[inst1]{organization={Princeton University},%Department and Organization
            city={Princeton},
            postcode={08544}, 
            state={New Jersey},
            country={USA}}
\affiliation[inst2]{organization={University of Maryland},%Department and Organization
        city={College Park, MD},
        postcode={20742}, 
        state={Maryland},
        country={USA}}
\affiliation[inst4]{organization={Princeton Plasma Physics Laboratory},%Department and Organization
            city={Princeton},
            postcode={08543}, 
            state={New Jersey},
            country={USA}}
\affiliation[inst3]{organization={Thea Energy},%Department and Organization
            city={Kearny},
            postcode={07032}, 
            state={New Jersey},
            country={USA}}

\begin{abstract}
Zernike polynomials serve as an orthogonal basis on the unit disc, and have proven to be effective in optics simulations, astrophysics, and more recently in plasma simulations. Unlike Bessel functions, Zernike polynomials are inherently finite and smooth at the disc center (r=0), ensuring continuous differentiability along the axis. This property makes them particularly suitable for simulations, requiring no additional handling at the origin. We developed \texttt{ZERNIPAX}, an open-source Python package capable of utilizing CPU/GPUs, leveraging Google’s \texttt{JAX} package and available on GitHub as well as the Python software repository PyPI. Our implementation of the recursion relation between Jacobi polynomials significantly improves computation time compared to alternative methods by use of parallel computing while still performing more accurately for high-mode numbers.

\end{abstract}


% \begin{highlights}
% \end{highlights}

\begin{keyword}
Zernike polynomials \sep Optics \sep Astrophysics \sep Spectral Simulations \sep Python \sep \texttt{JAX} \sep CPU/GPU Computing
\end{keyword}

\end{frontmatter}


\section{Introduction}
\label{sec:introduction}

In computational science and engineering, the use of orthogonal polynomials is essential for efficient numerical simulations and data analysis. Among these, Zernike polynomials have garnered significant attention due to their unique properties and versatility \cite{niu_zernike_2022}. Zernike polynomials, first introduced by Frits Zernike \cite{zernike_beugungstheorie_1934, zernike_diffraction_1934}, form an orthogonal basis on the unit disc, a characteristic that has proven invaluable in various domains, including optics simulations\cite{schwiegerling_representation_1995,schwiegerling_using_1997,iskander_optimal_2001, rodrigues_modeling_2019, klyce_advantages_2004, wan_analytic_2024}, astrophysics\cite{gravity_collaboration_improved_2021, capalbo_three_2021, bara_zernike_2014, bara_zernike_2015}, and more recently, plasma simulations of toroidal magnetic confinement machines \cite{desc, hudson_non-axisymmetric_2011, qu_coordinate_2020}. Unlike Bessel functions, which can exhibit singularities at the disc center, Zernike polynomials remain finite and smooth at the disc center (r=0), ensuring continuous differentiability along the axis.

In optics, Zernike polynomials are widely used to describe wavefronts in optical systems \cite{schwiegerling_representation_1995,schwiegerling_using_1997,iskander_optimal_2001, rodrigues_modeling_2019, klyce_advantages_2004, wan_analytic_2024}. They are particularly effective in representing aberrations in lenses and mirrors, aiding in the design and optimization of high-precision optical instruments. The benefits of Zernike polynomials in optics make them useful in astrophysics applications that include the analysis and simulation of telescope optics and the characterization of optical system performance \cite{gravity_collaboration_improved_2021, capalbo_three_2021, bara_zernike_2014, bara_zernike_2015}. They are employed to model the surface irregularities of telescope mirrors, enabling researchers to predict and mitigate the impact of these irregularities on image quality. Due to their orthogonality, numerical simulations making use of the spectral method to solve partial differential equations (PDEs) use Zernike polynomials as the basis functions \cite{boyd_comparing_2011}. Within plasma simulations \cite{desc, hudson_non-axisymmetric_2011, qu_coordinate_2020}, Zernike polynomials are used as a spectral basis to represent the toroidal cross-section of highly complex torus geometries of stellarators and tokamaks, hence facilitating solving the PDEs describing the plasma. The reach of Zernike polynomials extends far beyond the few applications mentioned here, with their utility spanning a wide range of fields in science and engineering \cite{niu_zernike_2022}.

The computational study of Zernike polynomials has a rich history, characterized by the development of numerous methods to enhance their evaluation and application. Initially, computational approaches relied on direct evaluation techniques. Despite their straightforward nature, these techniques were often computationally expensive and prone to numerical instability. Over time, more sophisticated methods have emerged, such as the use of recurrence relations \cite{chong_comparative_2003, novak_efficient_2013, singh_fast_2010, deng_stable_2016, papakostas_numerical_2008, kintner_recurrence_1976, prata_algorithm_1989}. These studies successfully addressed instability issues for higher mode numbers and provided fast algorithms for calculating Zernike polynomials. However, the proposed recurrence relations require additional formulas for edge cases, adding minor complexity to coding implementations, particularly on Graphics Processing Units (GPUs), which are relatively slow for conditional executions.

Recent advancements in parallel computing have spurred further efforts to utilize computational resources more efficiently. Qin et al. employed the Jacobi recursion relation in conjunction with the relationship between Zernike and Jacobi polynomials to decouple radial polynomials and vectorize operations \cite{qin_parallel_2012}. They reported faster execution times than previous recursive methods such as those by Prata, Kintner, and others. However, their implementation was limited to Central Processing Units (CPUs). Janssen et al. described Zernike polynomials in terms of the discrete Fourier transform (DFT) \cite{janssen_computing_2007}, and a subsequent study implemented the DFT approach for Zernike polynomial calculation on GPUs \cite{al-rawi_ultra-fast_2023}. Zhang et al. developed an algorithm to reduce memory usage and increase speed by using a balanced binary scheme \cite{zhang_balanced_2023}. Despite significant progress in improving the accuracy and speed of Zernike polynomial computations, there remains a notable lack of open-source packages that implement these advanced methods.

The widespread application of Zernike polynomials in the aforementioned scientific fields underscores the need for robust computational tools that can handle complex calculations efficiently. Addressing this demand, we introduce \texttt{ZERNIPAX}, an open-source Python package designed to exploit the computational power of both CPUs and GPUs. By leveraging Google’s \texttt{JAX} library \cite{jax2018github}, \texttt{ZERNIPAX} provides significantly faster computation and higher accuracy, particularly through the implementation of the Jacobi polynomial expression of Zernike polynomials and the recursion relation for Jacobi polynomials.

This paper presents \texttt{ZERNIPAX}'s capabilities and demonstrates its potential as a valuable tool for the scientific community. Whether in astrophysics, optics, or any field that relies on Zernike polynomials, \texttt{ZERNIPAX} offers a streamlined and powerful solution for researchers seeking to improve their computational workflows.

% - - - - - - - - - - - - - - - - - - - - - - - - - - - - - - - - - - 
\section{Zernike Polynomials}
Zernike polynomials are defined for $\rho \in [0,1]$ and $\theta \in [0,2\pi]$ as,
\begin{equation}
    \mathcal{Z}_n^m(\rho,\theta) = \begin{cases}\begin{aligned}
            &\mathcal{R}_n^{m}(\rho) \cos(m\theta) & \text{for } m\geq 0 \\[.3cm]
            &\mathcal{R}_n^{|m|}(\rho) \sin(|m|\theta) & \text{for } m < 0
        \end{aligned}
    \end{cases}
    \label{zernike-poly}
\end{equation}
where $\mathcal{R}_n^m$ is the radial part of the Zernike polynomials and it is defined as\footnote{For the derivation of the given formula, readers can refer to Appendix VII of \cite{born_principles_1999}.},
\begin{equation}
    \mathcal{R}_n^{|m|} (\rho) = \mathlarger{\mathlarger{\sum}}_{s=0}^{(n-|m|)/2} \frac{(-1)^s(n-s)!}{ s!\left( \cfrac{n+|m|}{2} - s\right)! \left( \cfrac{n-|m|}{2} - s\right)!  }\hspace{0.1cm} \rho^{n-2s} \label{radial-part-eq}
\end{equation}
which can also be written as
\begin{equation}
    \mathcal{R}_n^{|m|} (\rho) = \mathlarger{\mathlarger{\sum}}_{s=0}^{(n-|m|)/2} (-1)^s \binom{n-s}{s} \binom{n-2s}{\frac{n-|m|}{2}-s}\hspace{0.1cm} \rho^{n-2s} \label{radial-part-eq-binom}
\end{equation}

The mode numbers $n$ and $m$ are defined such that $n\geq0$, $|m|\leq n$, and the difference $n-|m|$ must be even. Therefore, possible values for m are $m= \{(-n), (-n+2), ... , (n-2), n\}$. Moreover, since binomial coefficients in Equation \ref{radial-part-eq-binom} are always integers, the coefficients of the Zernike polynomials are also integers. The value at the center of the unit disc can be found by setting $n-2s$ in Equation \ref{radial-part-eq} to 0 since it gives the only polynomial term that won't be canceled by $\rho$=0. So, substituting $s=n/2$ results in,
\begin{equation}
    \mathcal{R}_n^{|m|} (\rho=0) = \cfrac{(-1)^{n/2}}{ (-m/2)! (m/2)! }
\end{equation}
Since negative values for the factorial are not allowed, the only possible value for $m$ is 0. Notice that this removes the dependency on $\theta$ for $\rho$=0, aligning with the requirements for Fourier expansions of analytic functions on the polar plane, as outlined in \cite{lewis_physical_1990}. The properties of Zernike polynomials inherently meet these conditions, making them a strong choice as a basis for numerical simulations in polar coordinates.
\begin{equation}
    \mathcal{Z}_n^m(\rho=0,\theta) = \mathcal{R}_n^0(0) = \begin{cases}
        1  & \text{$n=4k$ where $k=0,1,2,3 ...$}\\
        -1  & \text{$n=4k-2$ where $k=0,1,2,3 ...$}\\
        0 & \text{otherwise}\\
    \end{cases} \label{zernike-at0}
\end{equation}

The $(-1)^s$ term in Equation \ref{radial-part-eq} introduces alternating signs in the polynomial terms, reducing the accuracy of floating-point summation and causing loss of significant digits. Additionally, the exponential term in $\rho$ results in the summation of terms with vastly different magnitudes, such as $10^{-1}$ and $10^{-20}$, which exacerbates round-off errors. Consequently, direct evaluation of Equation \ref{radial-part-eq} becomes numerically unstable. To mitigate these instabilities, recursion relations are commonly used as an alternative to direct polynomial evaluation. The recursion relation for Zernike polynomials can be derived as follows:
\begin{equation}
    \mathcal{R}_n^m(\rho) = \rho\left[ \mathcal{R}_{n-1}^{|m-1|}(\rho) + \mathcal{R}_{n-1}^{m+1}(\rho) \right] - \mathcal{R}_{n-2}^m(\rho) \label{zernike-recursion}
\end{equation}
The implementation requires special care for edge cases where $m+1>n-1$. This corresponds to $n=m$ and those polynomials are trivial to calculate with a single term $\mathcal{R}_{n}^{n}(\rho)=\rho^n$, however, this would require conditionals which will slow down the algorithm when implemented in \texttt{JAX}.

The radial part of the Zernike polynomials can also be calculated using Jacobi polynomials. The reader can refer to the \ref{appendix-jac} for proof of this relation.
\begin{equation}
    \mathcal{R}_n^m(\rho) = (-1)^{(n-m)/2} \rho^m  P_{(n-m)/2}^{m, 0} (1 - 2 \rho^2) \label{jacobi-relation}
\end{equation}
where $P_{n}^{\alpha, \beta}(x)$ is a Jacobi polynomial that is defined as,
\begin{equation}
    P_{n}^{\alpha, \beta}(x) = \sum_{s=0}^{n} \binom{n+\alpha}{n-s}  \binom{n+\beta}{s}  \left( \frac{x-1}{2} \right)^s \left( \frac{x+1}{2} \right)^{n-s} \label{jacobi}
\end{equation}

This allows numerical simulations to use stable recurrence relations for the Jacobi polynomials \cite{camacho-bello_high-precision_2014}. Here is the recursion relations for Jacobi polynomials,
\begin{multline}
    2n(c-n)(c-2)P_{n}^{\alpha,\beta}(\rho) = (c-1)[c(c-2)\rho + (a-b)(c-2n)]P_{n-1}^{\alpha,\beta}(\rho) \\
    - 2(a-1)(b-1)cP_{n-2}^{\alpha,\beta}(\rho) \label{jacobi-recursion}
\end{multline}
where,
\begin{equation}
    c = 2n + \alpha + \beta, \hspace{1cm} a = n +\alpha, \hspace{1cm} b = n + \beta
\end{equation}
Setting $\beta=0$ for Zernike polynomials and expanding the terms, one can get,
\begin{multline}
    P_{n}^{\alpha,0}(\rho) = \cfrac{(2n + \alpha-1)[(2n + \alpha)(2n + \alpha-2)\rho + \alpha^2]}{2n(n + \alpha)(2n + \alpha-2)} P_{n-1}^{\alpha,0}(\rho) \\
    - \cfrac{2(n +\alpha-1)(n-1)(2n + \alpha)}{2n(n + \alpha)(2n + \alpha-2)} P_{n-2}^{\alpha,0}(\rho)
\end{multline}
To assess the stability of this recursion, note that both the numerator and denominator in the coefficients of $P_{n-2}^{\alpha,0}(\rho)$ and $P_{n-1}^{\alpha,0}(\rho)$ are third-order polynomials in $n$. This structure implies that, as $n\rightarrow\infty$, these coefficients approach constant values, ensuring that errors introduced during each recursive step do not grow exponentially. Instead, the error growth rate stabilizes, reaching an asymptotic limit as $n$ increases. This characteristic is essential for maintaining numerical stability in high-mode calculations.
    
Derivatives of Jacobi polynomials can be calculated using the Jacobi polynomials itself with the following relation,
\begin{equation}
    \frac{d^k}{dx^k} P_n^{(\alpha, \beta)}(x) = \frac{\Gamma(\alpha + \beta + n + 1 + k)}{2^k\Gamma(\alpha + \beta + n + 1)} P_{n-k}^{(\alpha + k, \beta + k)}(x)
\end{equation}
Hence, the stable recursion relation can be incorporated for the derivatives too. However, instead of using the Gamma function for the relation, since for Zernike polynomials mode numbers $n$ and $m$ as well as $(n-m)/2$ are always integers, the argument of the Gamma function is an integer and one can use the factorial version to get more accurate results by allowing analytic cancellation of the factorial in the denominator.
\begin{equation}
    \frac{d^k}{dx^k} P_n^{(\alpha, \beta)}(x) = \frac{(\alpha+\beta+n+1)_k}{2^k} P_{n-k}^{(\alpha + k, \beta + k)}(x)
\end{equation}
where the Pochhammer function is defined as,
\begin{equation}
    (\alpha+\beta+n+1)_k = (\alpha+\beta+n+1)(\alpha+\beta+n+2)\dots(\alpha+\beta+n+k)
\end{equation}

For the implementation of Zernike polynomials, we choose the Jacobi form given in Equation \ref{jacobi-relation} along with Jacobi recursion relation expressed in Equation \ref{jacobi-recursion} since it is better suited to be implemented in \texttt{JAX} and offers higher numerical accuracy due to being a recursive formula.

% - - - - - - - - - - - - - - - - - - - - - - - - - - - - - - - - - - 
\section{Methodology}
\label{sec:Methodology}
In this section, we present the implementation of Zernike polynomials using the Python \texttt{JAX} package and the Jacobi recursion relationship for the polynomial evaluation. Unlike previous studies that focus on the calculation of a single mode, we optimized our CPU implementation to calculate a given set of modes as fast and as accurately as possible by storing the intermediate results of the recursion operations. In addition, we created an open-source Python library, \texttt{ZERNIPAX}, that includes different versions of the function which are optimized for different input cases such as a unique set of modes, inputs with duplicated modes
\footnote{For numerical simulations in 3D, a spectral code can use Zernike polynomials for the polar plane and another basis for the remaining dimension, such as Fourier basis in case of periodic boundaries (i.e. $f(r,\theta, \phi)=\sum_{lmn} \mathcal{R}_n^m(r)\mathcal{F}_m(\theta)\mathcal{F}_l(\phi)$ where $\mathcal{R}_n^m$ is the radial part of Zernike polynomials and $\mathcal{F}$ is the Fourier basis). The coupling of different bases would result in modes with the same $(n,m)$ but different $l$. We call these duplicated modes since the output of the Zernike polynomial function will be the same.}
, as well as CPU and GPU-optimized functions. Moreover, we have written our function to be compatible with forward and reverse-mode automatic differentiation such that users can take full advantage of \texttt{JAX} in projects involving Jacobian/Hessian calculations, etc.

\begin{algorithm}[H]
\caption{Algorithm to Compute Radial Part of Zernike polynomial on CPU}\label{cpu-algo}
\DontPrintSemicolon
\KwIn{$\bar{n}$, $\bar{m}$ (arrays of integers), $\bar{r}$ (array where each element $\in [0,1]$)}
\KwOut{2D array with dimensions based on $\bar{r}$ and $\bar{n}$}
Initialize output array\;
$\alpha_{max} \gets \max(|\bar{m}|)$\;
$r_{jacobi} \gets 1-2\bar{r}^2$\;
$\beta \gets 0$\;
\For{$\alpha = 0$ \KwTo $\alpha_{max}$}{
    Find $N_{max}$ for $\alpha$\;
    Compute $P^{\alpha,\beta}_{1}$ and $P^{\alpha,\beta}_{0}$ using Equation \ref{jacobi}\;
    \For{$n_{jacobi} = 2$ \KwTo $N_{max}$}{
        Compute $P^{\alpha,\beta}_{n_{jacobi}}$ from $P^{\alpha,\beta}_{n_{jacobi}-2}$, $P^{\alpha,\beta}_{n_{jacobi}-1}$ using Equation \ref{jacobi-recursion}\;
        Calculate $\mathcal{R}_{n}^{m}$ using Equation \ref{jacobi-relation} where $n=2n_{jacobi}+\alpha$\;
        \If{$n \in \bar{n}$ and $m \in \bar{m}$}{
            Store $\mathcal{R}_{n}^{m}$ in the correct column of output array
        }
        $P^{\alpha,\beta}_{n_{jacobi}-2} \gets P^{\alpha,\beta}_{n_{jacobi}-1}$\;
        $P^{\alpha,\beta}_{n_{jacobi}-1} \gets P^{\alpha,\beta}_{n_{jacobi}}$\;
    }
}
\end{algorithm}

The algorithm used in the CPU version is summarized in  Algorithm \ref{cpu-algo}. It can be observed from the recursion relation in Equation \ref{jacobi-recursion} that starting from known values $P_0^{\alpha,0}(x)$ and $P_1^{\alpha,0}(x)$, one can calculate $P_{(n-\alpha)/2}^{\alpha,0}$. Therefore, for a single $\alpha$ of Jacobi (or equivalently, $m$ of Zernike polynomial), calculation of highest $(n-\alpha)/2$ (or highest value of $n_{jacobi}$ referred to as $N_{max}$ in Algorithm \ref{cpu-algo}) includes the calculation of every lower value. So, a set of Zernike modes can be calculated with 2 nested for loops, the former for distinct $m$ values, and the latter for increasing $n$ values for the given $m$. Since we are calculating for a set of modes instead of a single one, our algorithm significantly minimizes the number of repeated calculations. A naive implementation of this algorithm might be very slow due to the lack of performance of Python built-in for-loops. However, we use \texttt{JAX} that supports parallelized for-loops and just-in-time (JIT) compilation to execute operations in XLA (Accelerated Linear Algebra) and thus are able to compute each mode extremely fast. Moreover, the stability of the Jacobi recursion relation helps us to compute higher-order polynomials accurately.

The method described above was implemented on a CPU and performed very well (to be quantified later in Section \ref{sec:results}). However, we observed significant performance degradation when running the same algorithm on a GPU. This slowdown was primarily due to interruptions in the loops for storing intermediate results and checking if these results were needed, particularly in lines 11 to 13 of Algorithm \ref{cpu-algo}. It has been seen that with \texttt{JAX}, the execution of for-loops on GPU is quite slow compared to CPU. Since this part of the algorithm uses a type of for-loop, the overall nested for-loop count is 3 and thus the GPU performance is greatly impacted. To address this in the function optimized for GPU, we employed a naive implementation and vectorized the execution of each mode to a separate GPU core as shown in Algorithm \ref{gpu-algo}. Each GPU core processes a mode number $(n,m)$ of the given set of $\bar{n}, \bar{m}$ input pairs and executes the Jacobi recursion relation using Equation \ref{jacobi-recursion}, without storing intermediate values or performing checks. The vectorization over each input pair reduces the number of for-loops to one. However, this approach introduces some redundancy; for example, if both $(12, 8)$ and $(12, 12)$ modes are requested, the computation for $(12, 12)$ will repeat calculations for $(12, 8)$. Despite these inefficiencies, the GPU-optimized implementation achieved faster computation times than the CPU-optimized version for higher-order modes and larger numbers of radial points. The parallelization and reduced checks effectively offset the additional computation time from redundant calculations.

The implementation of this algorithm in \texttt{JAX} requires careful design choices due to certain limitations imposed by JIT compilation. In particular, \texttt{JAX}'s JIT cannot handle functions where the output array dimensions depend on the specific values within the input data—only on the shapes of the input arrays. For example, identifying pairs $(n, m)$ where $n = n^*$ would yield output dimensions that vary based on input values, which is currently unsupported by JIT. To work within this constraint, we structured the CPU version of the algorithm to compute the complete set of modes, with $n$ values ranging from 0 up to the maximum $n$ specified and all feasible $m$ values for each $n$. Additionally, variable loop limits in the algorithm require implementing a custom derivative rule to ensure compatibility with \texttt{JAX}'s automatic differentiation system. These limitations are less pronounced in the GPU version, which can process each $(n, m)$ mode individually. However, for the CPU version, this design choice may affect performance in applications where only a single mode calculation is required, as it’s optimized for handling the full mode set. Although this limitation is noted here, it will not appear in the results, as this paper focuses on improving the calculation of a full set of modes rather than individual mode computation.

While we refer to these versions as CPU and GPU algorithms, both methods can technically run on either hardware, their name just reflects which hardware they were optimized to perform best on. Thus, users can flexibly use either algorithm according to their specific needs, regardless of the hardware used.

\begin{algorithm}
\caption{Algorithm to Compute Radial Part of Zernike polynomial on GPU}\label{gpu-algo}
\DontPrintSemicolon
\SetKwProg{vectorize}{do in parallel}{:}{}
\KwIn{$\bar{n}$, $\bar{m}$ (arrays of integers), $\bar{r}$ (array where each element $\in [0,1]$)}
\KwOut{2D array with dimensions based on $\bar{r}$ and $\bar{n}$}
$r_{jacobi} \gets 1-2\bar{r}^2$\;
$\beta \gets 0$\;
\vectorize{ for each $(n,m)$ in $(\bar{n},\bar{m})$}{
    $\alpha \gets m$\;
    Compute $P^{\alpha,\beta}_{1}$ and $P^{\alpha,\beta}_{0}$ using Equation \ref{jacobi}\;
    \For{$n_{jacobi} = 2$ \KwTo $n$}{
        Compute $P^{\alpha,\beta}_{n_{jacobi}}$ from $P^{\alpha,\beta}_{n_{jacobi}-2}$, $P^{\alpha,\beta}_{n_{jacobi}-1}$ using Equation \ref{jacobi-recursion}\;
        Calculate $\mathcal{R}_{n'}^{m}$ using Equation \ref{jacobi-relation} where $n'=2n_{jacobi}+\alpha$\;
        $P^{\alpha,\beta}_{n_{jacobi}-2} \gets P^{\alpha,\beta}_{n_{jacobi}-1}$\;
        $P^{\alpha,\beta}_{n_{jacobi}-1} \gets P^{\alpha,\beta}_{n_{jacobi}}$\;
    }
    Store $\mathcal{R}_{n}^{m}$
}
\end{algorithm}

% - - - - - - - - - - - - - - - - - - - - - - - - - - - - - - - - - - 
\section{Results}\label{sec:results}
First, we will show the accuracy of our implementation with Jacobi recursion by comparing it to the high-precision calculation of the standard form of Zernike polynomials given in Equation \ref{radial-part-eq}. The high precision method used to evaluate Equation \ref{radial-part-eq} includes calculating the coefficients first, and since every operation involves integers as operands and integers as output, the coefficients can be calculated exactly in Python which provides infinite precision for integers and integer arithmetics. The only operation that might cause floats is the division, however, one can easily deduce that from Equation \ref{radial-part-eq-binom}, all Zernike radial coefficients are integers. Hence, integer division can be used. Then, we evaluate the polynomial with these coefficients. To do so, we use \texttt{mpmath} \cite{mpmath}, a Python package allowing users to set arbitrary precision, and we evaluate polynomials up to order $n=m=50$ with the precision set to 100 decimal points which is equivalent to having 336-bit precision. This level of precision requires approximately 5–6 times the memory of a 64-bit representation, but due to additional overhead from arbitrary precision arithmetics, the actual memory usage exceeds six times that of standard 64-bit precision. The result with \texttt{mpmath} 100 decimal point precision is taken to be the exact values (see \ref{appendix-mpmath} for the validity). Due to the additional precision, this method is accurate but around 300 times slower compared to the same evaluation using 64-bit \texttt{numpy}. It is worth noting that by default \texttt{JAX} operations are in 32-bit precision but for the results shown in this paper, we set it to 64 bits. The results in this paper are obtained using Princeton University's Della cluster with a single core of a 2.8 GHz Intel Cascade Lake CPU and an 80GB memory Nvidia A100 GPU.

\begin{figure}
    \centering
    \includegraphics[width=\textwidth]{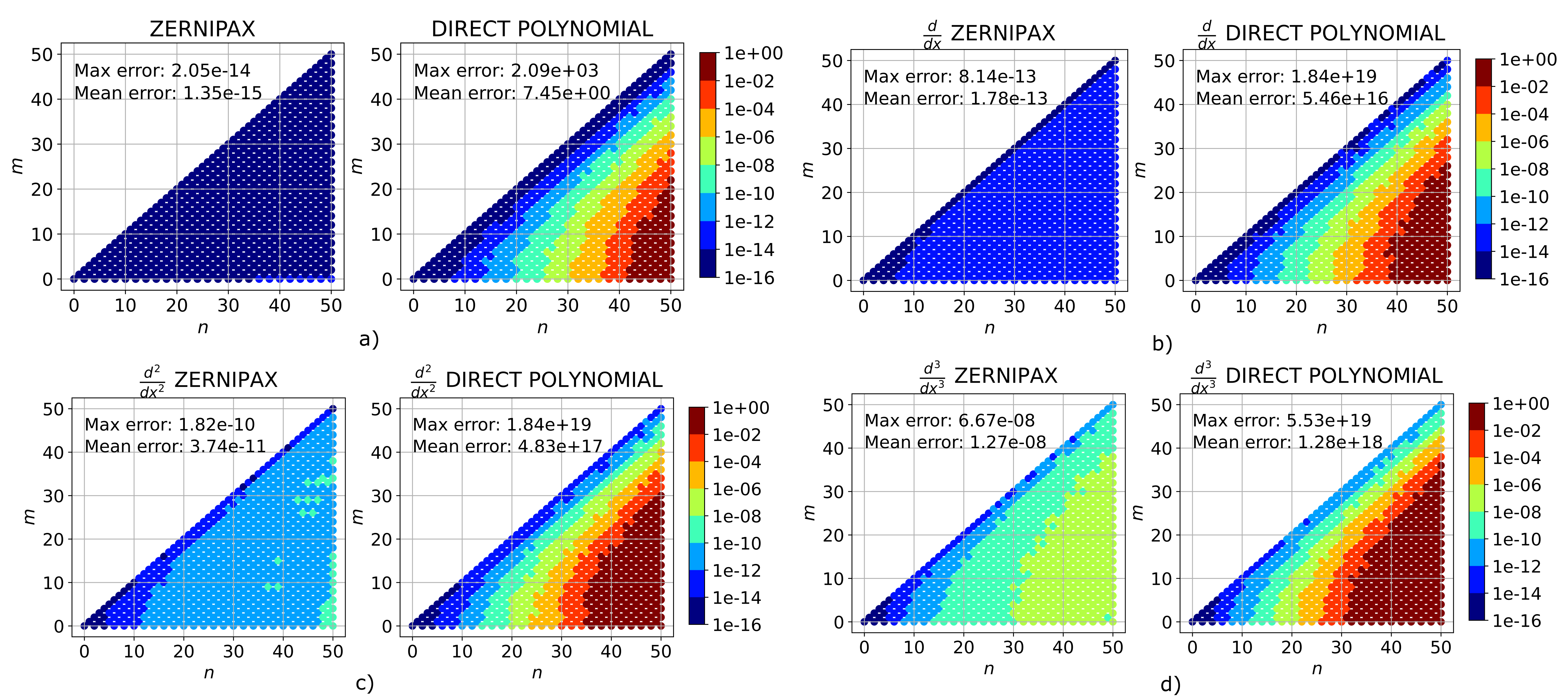}
    \caption{Accuracy of Jacobi recursion relation in \texttt{ZERNIPAX} (left) compared to direct polynomial evaluation (right) of \textbf{a)} radial part of Zernike polynomials \textbf{b)} first derivative \textbf{c)} second derivative \textbf{d)} third derivative. For the error, $\max_{x\in(0,1)}|Z_{nm}(x)-\Tilde{Z}_{nm}(x)|$, we evaluate both methods at 100 linearly spaced radial points for each $(n,m)$ mode in $n\in[0,50]$ and $m\in[0,50]$, and take the maximum absolute value difference with the exact calculation (with \texttt{mpmath} 100 significant digit precision), $\Tilde{Z}_{nm}$, for the error of that mode. The results, $Z_{nm}$, have been obtained using the CPU version of \texttt{ZERNIPAX} with 64-bit precision on a 2.8 GHz Intel Cascade Lake CPU.}
    \label{derivatives}
\end{figure}

\begin{figure}
    \centering
    \includegraphics[width=\textwidth]{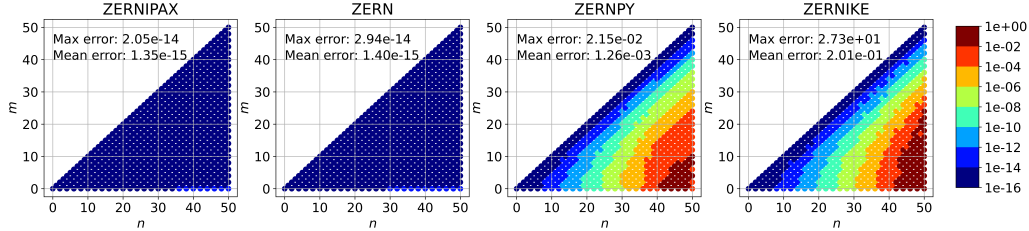}
    \caption{Comparison of the accuracies with 3 open-source codes, namely \texttt{ZERN}, \texttt{ZERNPY} and \texttt{ZERNIKE}, respectively. For the error, $\max_{x\in(0,1)}|Z_{nm}(x)-\Tilde{Z}_{nm}(x)|$, we evaluate every code at 100 linearly spaced radial points for each $(n,m)$ mode in $n\in[0,50]$ and $m\in[0,50]$, and take the maximum absolute value difference with the exact calculation (with \texttt{mpmath} 100 significant digit precision), $\Tilde{Z}_{nm}$, for the error of that mode. The results, $Z_{nm}$, have been obtained using a 2.8 GHz Intel Cascade Lake CPU with 64-bit precision.}
    \label{compare-all}
\end{figure}

In Figure \ref{derivatives}, it can be seen that Jacobi recursion is much more stable for higher mode numbers compared to the direct polynomial evaluation of Equation \ref{radial-part-eq} using 64-bit precision. We observe that although the polynomial coefficients can be calculated exactly, the direct polynomial evaluation in 64 bits causes floating point errors to accumulate for higher-order cases. Moreover, errors tend to increase for higher values of difference $n - m$, reaching a minimum along the $n = m$ line, and the maximum error occurs for the $n=50, m=0$ mode. As previously discussed, this instability arises from summing terms with vastly different magnitudes, which becomes particularly pronounced in higher-order polynomials in $\rho$ when using direct polynomial evaluation. In Figures \ref{derivatives} and \ref{compare-all}, the color bar range is set to be between $10^{-16}$ and 1 to provide sufficient resolution for both methods, ensuring a fair comparison without suggesting one is flawless, even though some error values are above 1. The actual maximum and average values are also indicated in the figures for clarity.

Unlike most of the previous studies on Zernike polynomials, in this paper, we are sharing our code as an open-source library, instead of only explaining the methodology. Therefore, we will also compare the performance of our code with the limited number of other open-source Zernike polynomial codes, such as \texttt{ZERN} \cite{zern-gh}, \texttt{ZERNIKE} \cite{zernike-gh} and \texttt{ZERNPY} \cite{zernpy-gh}, instead of comparing it to our own versions of different algorithms since the implementation could significantly affect the performance of the algorithm. In Figure \ref{compare-all}, it can be seen that the codes using Jacobi recursion relation, namely \texttt{ZERNIPAX} and \texttt{ZERN}, perform substantially better for high mode numbers, whereas libraries \texttt{ZERNIKE} (which uses direct polynomial evaluation), and \texttt{ZERNPY} (which uses direct polynomial evaluation until $n=10$ and then switches to the recursion relation of Zernike polynomials in Equation \ref{zernike-recursion}) suffer from numerical instabilities at higher mode numbers.

\begin{figure}
    \centering
    \includegraphics[width=\textwidth]{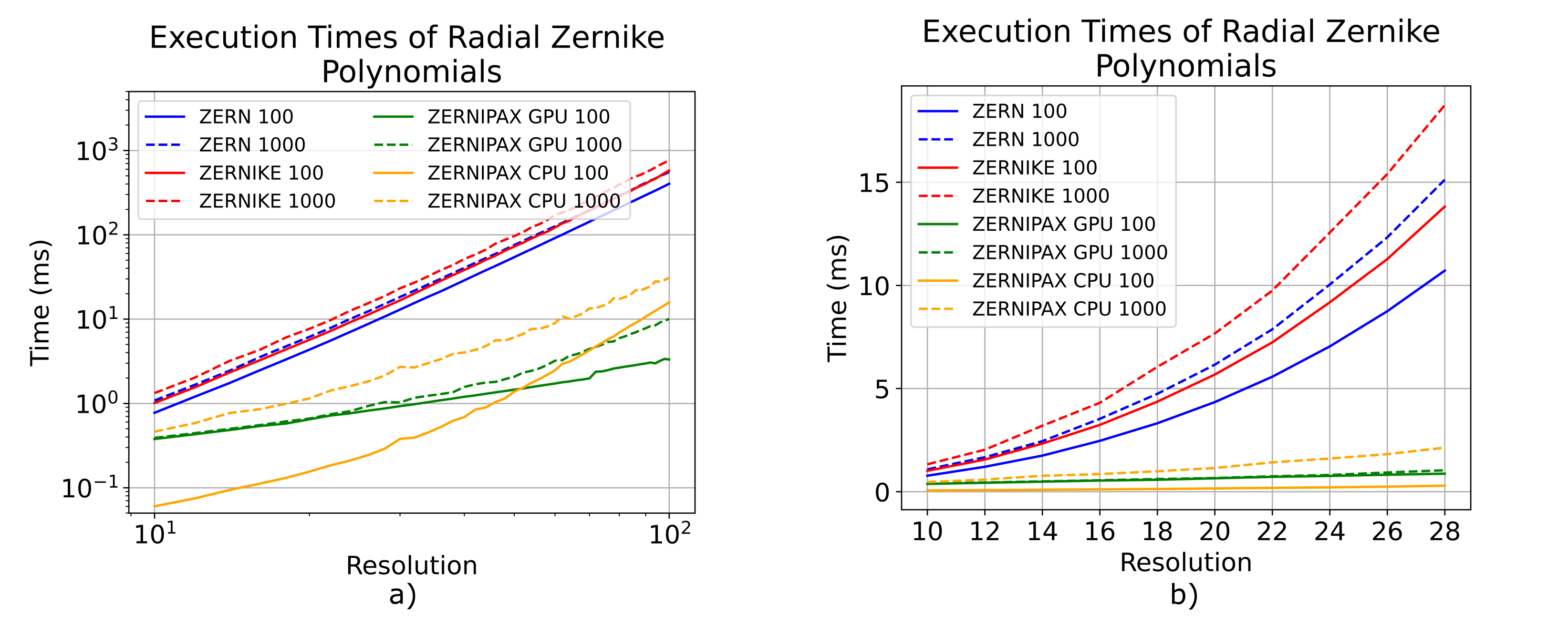}
    \caption{Execution time of the full set of radial Zernike polynomials for both 100 and 1000 linearly spaced points along radial direction \textbf{a)} for resolutions $n\in[10, 100]$ in log scale \textbf{b)} for resolutions $n\in[10,30]$ on a linear scale for packages \texttt{ZERN}, \texttt{ZERNIKE} and both CPU and GPU-optimized versions of \texttt{ZERNIPAX}. Packages \texttt{ZERN} and \texttt{ZERNIKE} can only use CPU resources. \texttt{ZERNIPAX} results don't include the compilation time which only affects the first run. The results have been obtained using a 2.8 GHz Intel Cascade Lake CPU and 80GB memory Nvidia A100 GPU with 64-bit precision.}
    \label{time-compare}
\end{figure}

To compare computation time, we calculate the radial part of Zernike polynomials at 100 and 1000 linearly spaced points of $\rho$ for a full set of Zernike polynomial modes for resolutions ranging from 10 to 100. Since \texttt{ZERNPY} is documented to be slow for mode numbers greater than 50, it is not included in the time comparison. In Figure \ref{time-compare}, the time comparison of three different packages for both radial resolutions is shown with CPU and GPU versions of \texttt{ZERNIPAX}. \texttt{ZERNIPAX} execution is more than an order of magnitude faster. In the GPU case, we see even faster computations with \texttt{ZERNIPAX} for higher modes. However, the other two packages perform the same, since they are not able to run on GPU. Moreover, it can be seen that for low resolutions, the CPU version of \texttt{ZERNIPAX} is faster than the GPU version for 100 radial points, this is due to the creation of kernels on GPU being overhead and the inefficiency of the GPU version mentioned in Section \ref{sec:Methodology}. As we increase the number of radial points, the gains due to parallelization become more pronounced, and thus the GPU version becomes faster than the CPU version, even for low resolutions.

An important point to note about the implementation with \texttt{JAX} is that there are $\sim$250 milliseconds of overhead due to the just-in-time compilation, which is a single-time cost paid at the first execution of the function that is not included for the results of Figure \ref{time-compare}. The compilation time is almost independent of the given resolution, for example, $n=10$ takes 240ms and $n=200$ takes 260ms to compile. While the overhead cost is significant compared to execution time for low-resolution cases, we observe that for mode numbers higher than 80, the execution times of the other two codes are longer even if we include the compilation time. Still, for the low resolutions, if the function is called with the same input dimensions
\footnote{The just-in-time compilation in \texttt{JAX} compiles and stores the compiled function in runtime for future use. The compiled version of the function is used if the input arguments have the same dimensions as one of the previously compiled functions. However, if the function is called with an unprecedented input shape, \texttt{JAX} will recompile the function. Notice that the values can be different, the important point is the shapes of the arrays.} 
multiple times, it can be said that our algorithm gives significantly faster results.

% - - - - - - - - - - - - - - - - - - - - - - - - - - - - - - - - - - 
\section{Conclusion}
In this study, we developed an efficient Python implementation of Zernike polynomials using \texttt{JAX} and the Jacobi recursion relation, achieving high computational accuracy even for very high mode numbers. Notably, the CPU version of our code is approximately ten times faster than comparable open-source implementations while maintaining a maximum error of only 2.05e-14 for modes up to $n = 50$.

We developed a separate GPU-optimized version of our implementation, achieving faster computation times for higher-order polynomials through parallel processing. GPU performance scales with the number of modes and radial points, as computation is vectorized in both dimensions. While the GPU excels at high resolutions, lower resolutions may still be faster on the CPU. This flexibility enables our implementation to support both low and high-fidelity simulations on various hardware.

Our results indicate that this faster and more precise implementation benefits a range of fields, including optics, astrophysics, plasma physics, and other areas that depend on Zernike polynomials for high-performance modeling. The increased computational efficiency not only enhances accuracy but also opens up possibilities for real-time applications, making this approach especially valuable in scenarios demanding both speed and precision with high resolutions.

The integration of \texttt{JAX} in our Python implementation provides both high-performance computation and ease of use, especially valuable for optimization routines requiring automatic differentiation for Jacobian calculations. This combination of efficiency and user-friendly code broadens the applicability of our implementation across scientific and engineering contexts. However, due to the JIT compilation overhead, our algorithm is most effective in applications requiring repeated calculations of multiple Zernike polynomials, with less computational savings for one-off, single-mode evaluations.

In summary, our work advances the computational evaluation of Zernike polynomials by combining the robustness of the Jacobi recursion with \texttt{JAX}’s efficiency. This tool can be readily applied in various fields, from optical analysis to image processing. Future work may explore further optimizations to broaden its scope and improve performance.

\section{Funding}
% the PPPL stell foundry grant 
This work is funded through the SciDAC program by the US Department of Energy, Oﬃce of Fusion Energy Science and Oﬃce of Advanced Scientific Computing Research under contract No. DE-AC02-09CH11466, as well as DE-SC0022005 and the Hidden Symmetries grant from the Simons Foundation (560651). The United States Government retains a non-exclusive, paid-up, irrevocable, world-wide license to publish or reproduce the published form of this manuscript, or allow others to do so, for United States Government purposes.

\section{Data availability statement}
The source code to generate the results and plots in this study are openly available on GitHub at \url{https://github.com/PlasmaControl/Zernike-Paper.git}. The latest version of \texttt{ZERNIPAX} can be accessed through \url{https://github.com/PlasmaControl/ZERNIPAX.git} or can be used in Python using pip package.

\section{Declaration of generative AI and AI-assisted technologies in the writing process}
During the preparation of this work, the author Yigit Gunsur Elmacioglu used OpenAI's ChatGPT in order to increase the readability of the text. After using this tool, the authors reviewed and edited the content as needed and take full responsibility for the content of the publication.

% \section{References}
\bibliographystyle{elsarticle-num} 
\bibliography{references-revised}





% \end{thebibliography}

\appendix

\section{Relation between Zernike polynomials and Jacobi polynomials}
\label{appendix-jac}
For the special case of $\beta=0$, $x=1-2\rho^2$ and $\alpha=m$, Jacobi polynomials can be written using Equation \ref{jacobi},
\begin{equation}
    P_{n}^{m, 0}(1-2\rho^2) = \sum_{s=0}^{n} (-1)^s\binom{n+m}{n-s} \binom{n}{s} \rho^{2s}(1-\rho^2)^{n-s} \label{eq_jac}
\end{equation}
Now use the binomial theorem to expand $(1-\rho^2)^{n-s}$,
\begin{equation}
    (1-\rho^2)^{n-s} = \sum_{k=0}^{n-s} (-1)^k \binom{n-s}{k} \rho^{2k}
\end{equation}
Substitute this in eq \ref{eq_jac},
\begin{equation}
    P_{n}^{m, 0}(1-2\rho^2) = \sum_{s=0}^{n} (-1)^s \binom{n}{s} \binom{n+m}{n-s}  \rho^{2s} \sum_{k=0}^{n-s} (-1)^k \binom{n-s}{k} \rho^{2k}
\end{equation}
Now, rearrange the terms,
\begin{equation}
    P_{n}^{m, 0}(1-2\rho^2) = \sum_{s=0}^{n}\sum_{k=0}^{n-s} (-1)^{(s+k)} \binom{n+m}{n-s} \binom{n}{s} \binom{n-s}{k} \rho^{2(s+k)}
\end{equation}
Substitute $j=s+k$, hence $k=j-s$ ,
\begin{align}
    P_{n}^{m, 0}(1-2\rho^2) &= \sum_{s=0}^{n}\sum_{j-s=0}^{j-s=n-s} (-1)^{j} \binom{n+m}{n-s} \binom{n}{s} \binom{n-s}{j-s}\rho^{2j}\\
    P_{n}^{m, 0}(1-2\rho^2) &= \sum_{s=0}^{n}\sum_{j=s}^{n} (-1)^{j} \binom{n+m}{n-s} \binom{n}{s} \binom{n-s}{j-s}\rho^{2j}  
\end{align}
Here, we can change the order of summation, it is better to use table to find new limits,
\[
\begin{array}{c|cccccc}
  & 0 & 1 & 2 & \cdots & n-1 & n \\
\hline
s = 0 & \times & \times & \times & \cdots & \times & \times \\
s = 1 &       & \times & \times & \cdots & \times & \times \\
s = 2 &       &       & \times & \cdots & \times & \times \\
\vdots &       &       &       & \ddots & \vdots & \vdots \\
s = n-1 &       &       &       &       & \times & \times \\
s = n &       &       &       &       &       & \times \\
\end{array}
\]
Each \(\times\) represents a valid pair \((s, j)\). This can be re-written in terms of summation over $j$ first, then $s$,
\[
\begin{array}{c|cccccc}
  & 0 & 1 & 2 & \cdots & n-1 & n \\
\hline
j = 0 & \times &       &       &       &       &       \\
j = 1 & \times & \times &       &       &       &       \\
j = 2 & \times & \times & \times &       &       &       \\
\vdots &       &       &       & \ddots &       &       \\
j = n-1 & \times & \times & \times & \cdots & \times &       \\
j = n & \times & \times & \times & \cdots & \times & \times \\
\end{array}
\]
Since the \((s, j)\) pairs are the same, the nested summation can be written as,
\begin{equation}
    P_{n}^{m, 0}(1-2\rho^2) = \sum_{j=0}^{n} \rho^{2j}   \sum_{s=0}^{j} (-1)^{j} \frac{(n+m)!}
    {(n-s)!(m+s)!}\frac{n!}{s!(n-s)!}\frac{(n-s)!}{(j-s)!(n-j)!}
\end{equation}
\begin{equation}
    P_{n}^{m, 0}(1-2\rho^2) = \sum_{j=0}^{n} \rho^{2j}   \sum_{s=0}^{j} (-1)^{j} \frac{(n+m)!}{(n-s)!(m+s)!} \frac{j!}{s!(j-s)!} \frac{n!}{j!(n-j)!} 
\end{equation}
\begin{equation}
    P_{n}^{m, 0}(1-2\rho^2) = \sum_{j=0}^{n} (-1)^{j} \binom{n}{j} \rho^{2j}   \sum_{s=0}^{j}  \binom{n+m}{n-s} \binom{j}{s}  \label{eq_jacobi2simplify}
\end{equation}
Now, we need to use a property of the binomial coefficients. Consider,
\begin{equation}
    (1+x)^n = \sum_{k=0}^{n} \binom{n}{k}x^k 
\end{equation}

\begin{equation}
    \begin{split}
        (1+x)^{n+m}(1+x)^j =&  \left(\sum_{k=0}^{n+m} \binom{n+m}{k}x^k\right) \left(\sum_{k=0}^{j} \binom{j}{k}x^k\right) \\
        =& \left(\sum_{s=-m}^{n} \binom{n+m}{n-s}x^{n-s}\right) \left(\sum_{k=0}^{j} \binom{j}{k}x^k\right) 
    \end{split}
\end{equation}  

\begin{equation}
    (1+x)^{n+m+j} = \sum_{k=0}^{n+m+j} \binom{n+m+j}{k}x^k 
\end{equation}
For $x^\gamma$ coefficient, we have 
\begin{equation}
    \binom{n+m+j}{\gamma} x^{\gamma}= \sum_{k=0}^{\gamma} \binom{j}{k} \binom{n+m}{\gamma-k} x^{\gamma}
\end{equation}
In previous step, I used $n-s+k=\gamma$ and $s=n+k-\gamma$, hence $n-s=\gamma-k$. Now, let's substitute $\gamma=n$,
\begin{equation}
    \binom{n+m+j}{n} = \sum_{k=0}^{j} \binom{n+m}{n-k} \binom{j}{k}
\end{equation}
We can finally use this relation to simplify eq \ref{eq_jacobi2simplify},
\begin{equation}
    P_{n}^{m, 0}(1-2\rho^2) = \sum_{j=0}^{n} (-1)^{j}  \rho^{2j}  \binom{n}{j} \binom{n+m+j}{n}
\end{equation}
Lets's multiply last equation by $\rho^m$ and $(-1)^n$,
\begin{equation}
    (-1)^n\rho^mP_{n}^{m, 0}(1-2\rho^2) = \sum_{j=0}^{n} (-1)^{j+n}  \rho^{2j+m} \binom{n}{j} \binom{n+m+j}{n}
\end{equation}
Substitute $j=n-s$,
\begin{equation}
    (-1)^n\rho^mP_{n}^{m, 0}(1-2\rho^2) = \sum_{n-s=0}^{n-s=n} (-1)^{2n-s}  \rho^{2n+m-s}  \binom{n}{n-s} \binom{2n+m-s}{n}
\end{equation}
\begin{equation}
    (-1)^{\frac{l-m}{2}}\rho^mP_{\frac{l-m}{2}}^{m, 0}(1-2\rho^2) = 
    \sum_{s=0}^{(l-m)/2} (-1)^{s}  \rho^{l-2s}  \binom{\frac{l-m}{2}}{s} \binom{l-s}{\frac{l-m}{2}}
\end{equation}
\begin{equation}
    (-1)^{\frac{l-m}{2}}\rho^mP_{\frac{l-m}{2}}^{m, 0}(1-2\rho^2) = 
    \sum_{s=0}^{(l-m)/2} (-1)^{s} \cfrac{\frac{l-m}{2}!}{s!(\frac{l-m}{2}-s)!} \cfrac{(l-s)!}{\frac{l-m}{2}!(\frac{l+m}{2}-s)!} \rho^{l-2s}  
\end{equation}
\begin{equation}
    \begin{split}
        \mathcal{R}_l^{m} (\rho) 
        &= (-1)^{\frac{l-m}{2}}\rho^mP_{\frac{l-m}{2}}^{m, 0}(1-2\rho^2) \\
        &= \mathlarger{\mathlarger{\sum}}_{s=0}^{(l-m)/2} \frac{(-1)^s(l-s)!}{ s!\left( \cfrac{l+m}{2} - s\right)! \left( \cfrac{l-m}{2} - s\right)!}  \hspace{0.1cm} \rho^{l-2s} 
    \end{split}
\end{equation}
which is exactly equivalent to the radial part of the Zernike polynomials.

\section{Validity of chosen precision in \texttt{mpmath}}
\label{appendix-mpmath}

\begin{figure}
    \centering
    \includegraphics[width=0.55\linewidth]{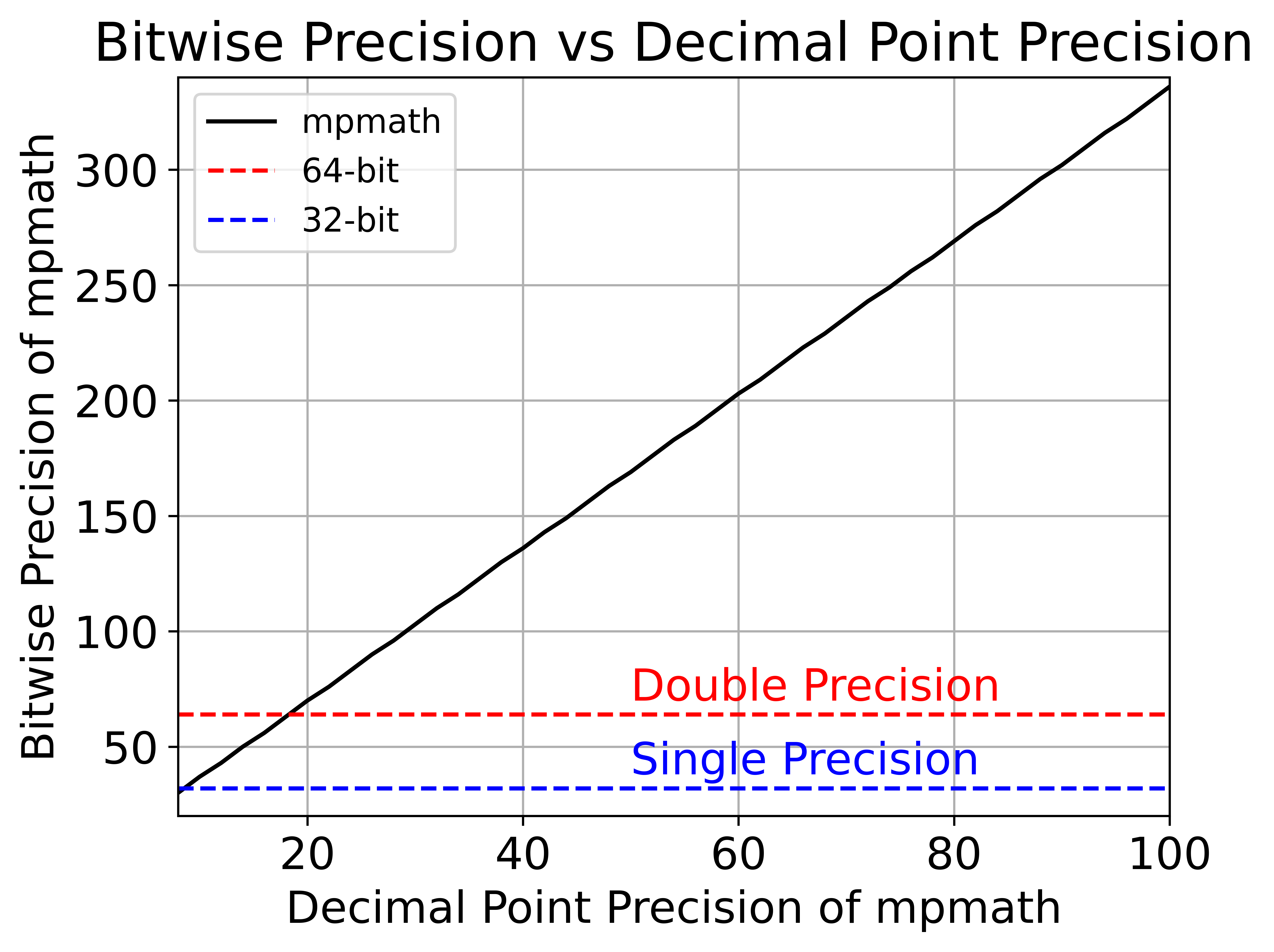}
    \caption{The corresponding bit precision of different \texttt{mpmath} decimal point precisions shown for range $[8,100]$. 64-bit precision can be achieved by setting \texttt{dps} to 18. The minimum value chosen for this figure \texttt{dps}=8 has 30-bit and the maximum value \texttt{dps}=100 has 336-bit precision.}
    \label{bitwise}
\end{figure}

The accuracy of different algorithms is tested using the \texttt{mpmath} package which allows arbitrary precision in Python. A user can set different decimal point precisions with the option \texttt{mpmath.mp.dps} that will change the number of significant digits used in \texttt{mpmath} functions such as \texttt{fsub()} for subtraction and \texttt{fadd()} for addition. The corresponding bit precision can be found using \texttt{mpmath.mp.prec} and the relation is shown in Figure \ref{bitwise}.

We performed the direct polynomial evaluation of Zernike polynomials up to $n,m$=100 with different decimal point precisions and compared the difference with an excessively high value of 200 significant digits. Since the machine precision of 64 bits is aimed for the final result, we use the native Python subtraction operation for error calculation.

\begin{figure}
    \centering
    \includegraphics[width=0.55\linewidth]{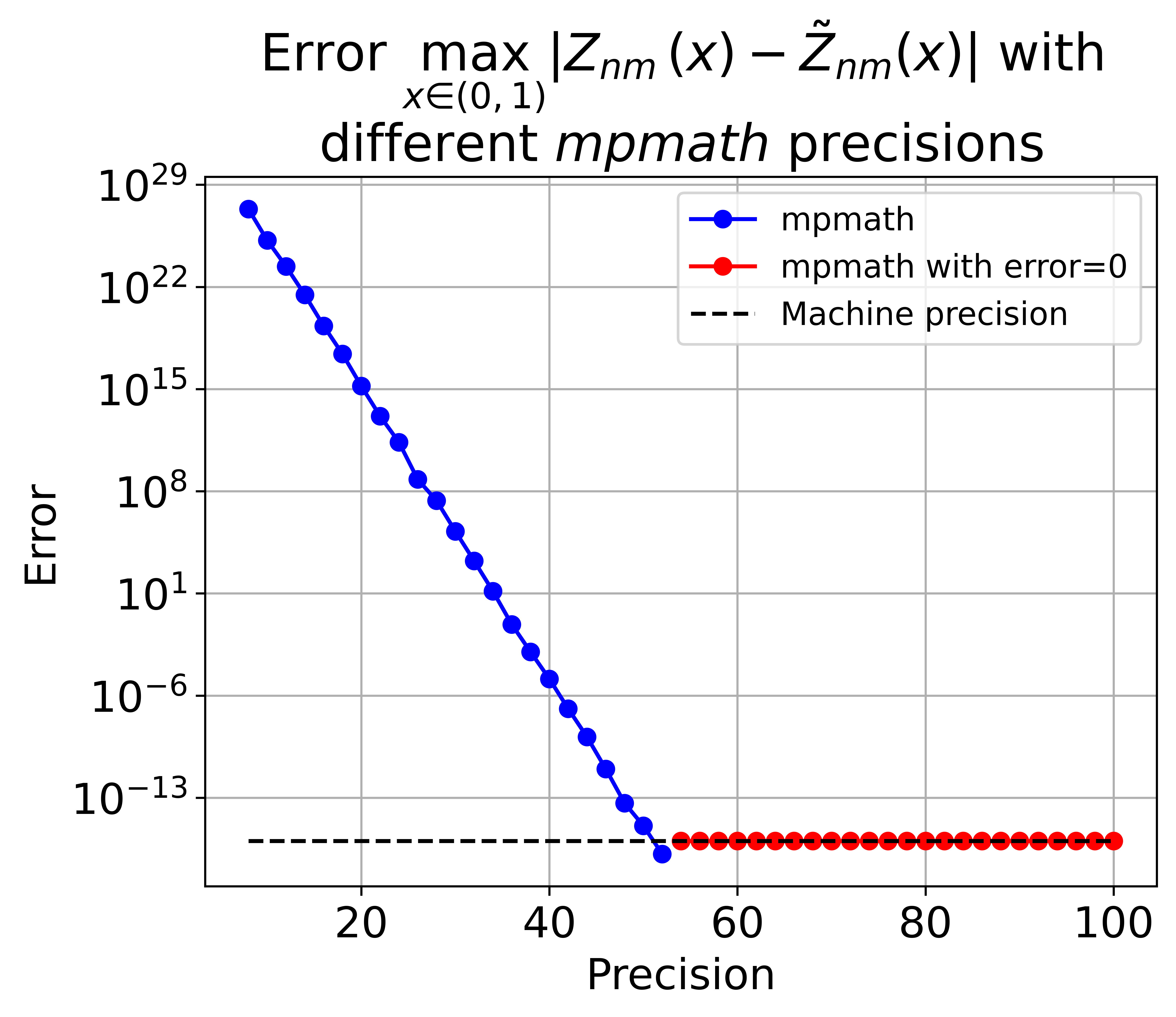}
    \caption{The maximum difference between Zernike polynomials up to $n,m$=100 calculated with lower \texttt{dps} and 200 \texttt{dps}, $\Tilde{Z}_{nm}(x)$. For the log scale, 0's are set to $2^{-53}$.}
    \label{prec-compare}
\end{figure}

After \texttt{mpmath.mp.dps}=54, the difference calculated with Python subtraction is 0. This corresponds to 183-bit precision. To be able to show the results on a log scale in Figure \ref{prec-compare}, we have set the 0 values to 64-bit machine precision that is $2^{-53}$. Although the required precision is 54 \texttt{dps}, we have chosen 100 \texttt{dps} for the results of this paper just to be on the safe side and to be able to use the same functions for higher mode numbers which require higher precision. This corresponds to 336-bit precision.

\end{document}